# A Gaussian mixture model based contrast enhancement

Mohsen Abdoli[1], Hossein Sarikhani[1], Mohammad Ghanbari[2, 3], and Patrice Brault[4]

Sharif University of Technology, Tehran, Iran[1],
University of Tehran, Tehran, Iran[2]
University of Essex, Colchester, UK[3]
CNRS, L2S Laboratory of Signals and Systems, Gif-sur-Yvette, France[4],
{abdoli, sarikhani}@ce.sharif.edu, ghan@essex.ac.uk, and patrice.brault@lss.supelec.fr

*Abstract* — In this paper, a method for enhancing low contrast images is proposed. This method, called Gaussian Mixture Model based Contrast Enhancement (GMMCE), brings into play the Gaussian mixture modeling of histograms to model the content of the images. Based on the fact that each homogeneous area in natural images has a Gaussian-shaped histogram, it decomposes the narrow histogram of low contrast images into a set of scaled and shifted Gaussians. The individual histograms are then stretched by increasing their variance parameters, and are diffused on the entire histogram by scattering their mean parameters, to build a broad version of the histogram. The number of Gaussians as well as their parameters are optimized to set up a GMM with lowest approximation error and highest similarity to the original histogram. Compared to the existing histogram-based methods, the experimental results show that the quality of GMMCE enhanced pictures are mostly consistent and outperform other benchmark methods. Additionally, the computational complexity analysis show that GMMCE is a low complexity method.

## I. INTRODUCTION

As the use of digital images grows in different applications, the need for more efficient methods of image enhancement to be applied on degraded images is perceived. One of the common degradations on images is the lack of contrast which can be due to the poor lighting conditions, such as extremely dark or bright environment [1]. The interpretation of low contrast in terms of histogram representation of digital images is that in images with low contrast content, the distribution of intensity values has low variance and consequently, their histograms have narrow shapes. The solutions to this problem, called contrast enhancement methods, have several applications in medical imaging [2], remote sensing [3], machine vision applications [4], consumer electronics [5] and so forth.

Though diverse classes of approaches have been proposed for this problem [6, 7], they can be classified into two general categories: histogram based and non-histogram based methods. Regarding different applications and their restrictions, one category may be preferred to the other.

One of the leading works on contrast enhancement is Histogram Equalization (HE) [8], where it tries to spread out the intensity values of the histogram on the entire intensity range. In other words, it effectively broadens out the narrow histogram of a low contrast image and generates its broadened version in such a way that the visual quality is improved.

Despite its simplicity, this straightforward method suffers from major drawbacks such as inability to preserve overall brightness of the image when the raw image is too dark or too bright [9] or over-enhancing the histogram when there are large peaks in the histogram [10]. To overcome these well-known drawbacks some extensions to HE have been proposed.

The first generation of extensions to HE are the brightness-preserving methods. They choose a specific intensity value from the dynamic range of the histogram to be fixed during the process of broadening the histogram [9, 11, 12, 13, and 14]. This intensity value acts as a separation point and consequently, the output histogram would consist of two sub-histograms, independently equalized with HE and shared a joint intensity value at their separation point.

Preserving a single intensity value of the histogram is not necessarily adequate to generate a high contrast and at the same time a visually fine image. Hence, the next generation of extensions aimed at preserving more than one intensity value of the histogram so that the histogram is separated more than once. These methods, called recursive brightness-preserving methods, perform similarly to the brightness-preserving methods except that they use a parameter named "Recursion Level (RL)" which indicates the number of separation points at which the current sub-histogram is split in two [15, 16].

All abovementioned derivatives of HE can be considered as "static" methods, since the number of selected separation points and their positions on the histogram before and after enhancement remain unchanged. As an alternative idea, two other sub-categories named "semi-dynamic" and "dynamic" methods have been introduced where the separation points and their positions are no longer predetermined. In the semi-dynamic methods, either the number of separation points is fixed, yet their positions might change during the contrast enhancement [17 – 19], or the positions of the separation points are fixed but their numbers is variable [20]. In the dynamic methods both the number of separation points and their positions can change depending on the characteristics of the image [21, 22].

As mentioned, there are also non-histogram based methods which improve the visual quality of images by taking advantage of features other than histogram characteristics. Generally, these methods need to have access to spatial attributes instead of global histogram characteristics. For instance, in order to enhance the contrast, a non-histogram based method might exploit over and under exposure versions of a low contrast image [23], statistical features of image related to stochastic resonance [24] or a genetic algorithm with non-histogram based chromosome structure [25]. Actually, there are several research works fitting in the non-histogram based category, but they are out of the scope of this study as they do not concern the proposed method in this paper.

In the present paper, a dynamic histogram-based contrast enhancement method is proposed. The main idea behind this method is based on the fact that in natural images, each homogeneous area is assumed to have a corresponding Gaussian-shaped region in the histogram of the image. Therefore, each histogram can be approximated with a combination of a limited number of Gaussian-shaped sub-histograms corresponding to its major regions. A greedy optimization method is proposed to obtain the optimum number of Gaussians as well as their parameters. Since mean parameters of the found Gaussians correspond to the average intensity value of major regions, they are assumed to be the most dominant intensity levels of the image. Consequently, the process of contrast enhancement is performed by spreading out these dominant intensity levels on the entire range of the histogram. This technique increases the contrast between the individual major regions of the image which improves the global contrast of the image.

The rest of the paper is organized as follows: in section II details of the histogram modeling as well as the proposed method are discussed. In section III, the experimental results plus some discussions on the relative performance of the proposed method against the other existing methods are presented and finally, section IV concludes the paper.

## II. GAUSSIAN MIXTURE MODEL BASED CONTRAST ENHANCEMENT (GMMCE)

In this section, first the idea behind using the Gaussian Mixture Modeling (GMM) to model the structure of a histogram is clarified and then the details of the proposed Gaussian mixture model based contrast enhancement method are explained.

### A. Histogram Modeling by Gaussians

Any arbitrary image can be assumed to be composed of individual meaningful regions occupying near- homogeneous areas of the image. Each region in natural images has a Gaussian-shaped histogram where the means of the Gaussian histograms indicate their corresponding average intensity levels and the variance corresponding to their texture details. These Gaussians are separated by their mean values and spread out with their variances, thus forming the global histogram [22]. Based on the fact that low contrast images have narrow histograms, if one departs the important means from each other, the contrasts of individual areas are enhanced and the visual quality of the image is improved.

In other words, the structure of an image is directly reflected in its histogram in a manner that any significant peak in the histogram is actually the mean intensity value corresponding either to a vast near-homogeneous zone of the image, or to several zones which together occupy a major portion of the area. In either case, these intensity levels are particularly important to the global visual quality of the image and should be carefully treated during any enhancement process.

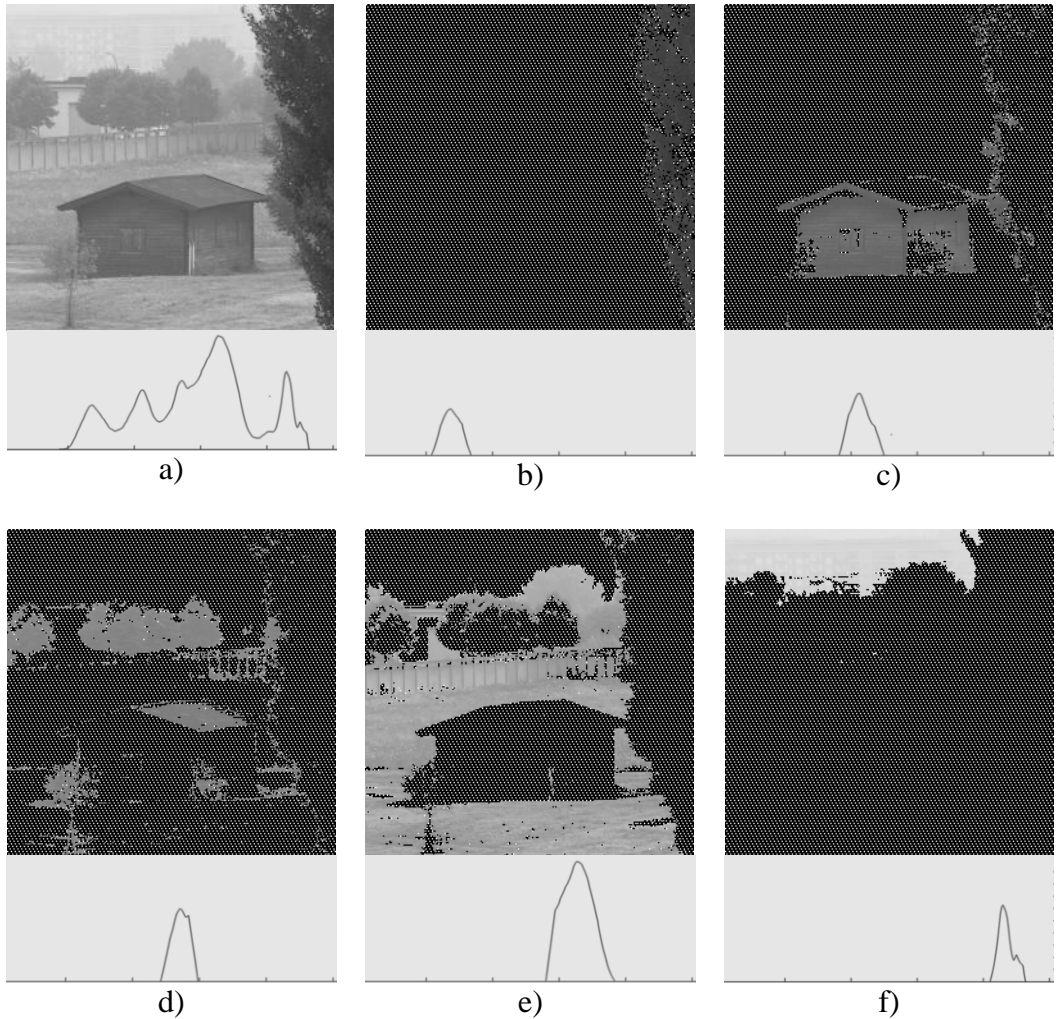

**Fig 1. a) A natural image and its global histogram. b) through f) 5 near-homogeneous regions and their independently calculated sub-histograms.**

Fig. 1 shows an image with five major near-homogeneous regions as well as their corresponding sub-histograms. Fig.1a is the original image with its histogram and Fig.1b through Fig.1f are the regions corresponding to the dominant intensity levels with their sub-histograms. Note that the dark backgrounds in all sub-images are related to the discarded areas that have been extracted from the original image and are not included in the sub-histograms. These images are generated by the proposed method which will be discussed in this section.

As can be seen, each meaningful area in the original image is explicitly related to a local peak in the global histogram. More importantly, the frequencies around these peaks form a Gaussian-shaped curve. Thus, it can be concluded that this observation can lead us to represent all natural images by Gaussian Mixture Modeling (GMM) where the number of Gaussians depends on the complexity of the content. A greedy algorithm is adopted to find the number of Gaussians as well as the parameters of their probability density functions which will be discussed later in this section.

*B. GMM Formulation*

Each arbitrary histogram can be formulated as a weighted sum of *k* Gaussians which requires to estimate three vectors of *k* parameters, namely mean $\mu$, variance $\sigma^2$ and scaling factor $\omega$. Thus the continuous GMM of the histogram can be expressed as:

$$h_{GMM}^{k}(I\mid\mu,\sigma^{2},\omega)=\sum_{j=1}^{k}\omega_{j}N(I\mid\mu_{j},\sigma_{j}^{2}) \qquad ,I=0,1,...,L-1 \qquad (1)$$

where $N(I\mid\mu_j,\sigma_j^2)$ is the $j^{th}$ Gaussian probability density function (PDF) at intensity level $I \in [0, L-1]$ with variance $\sigma_j^2$ and mean $\mu_i$ [26]. According to (1), the height of each Gaussian is scaled by $\omega_j$ which is proportional to the area is occupied by its corresponding region in the image. In other terms, the larger is a region, the more dominant is its corresponding Gaussian in the histogram, by having the highest peak in the global histogram.

Fig. 2 demonstrates the effect of increasing the number of Gaussians *k*, on the accuracy of the approximated histogram. As curves in Fig. 2d show, the absolute residual approximation error with 10 Gaussians (i.e. GMM(10) in Fig. 2c) is small, virtually making the resulting GMM accurate enough to be used in the enhancement process. However, increasing the number of Gaussians can be interpreted as increasing the number of dominant intensity levels. This has the drawback of having many Gaussians, as well as Gaussians with minor peaks, which represent a small fraction of the global intensity. The ultimate effect is that the increased number of Gaussians not only increases the complexity of image enhancement, but also has a negative effect on its quality. Since, having many Gaussians to decrease the absolute residual error, would cause considering non-dominant regions as important as the actual dominant regions.

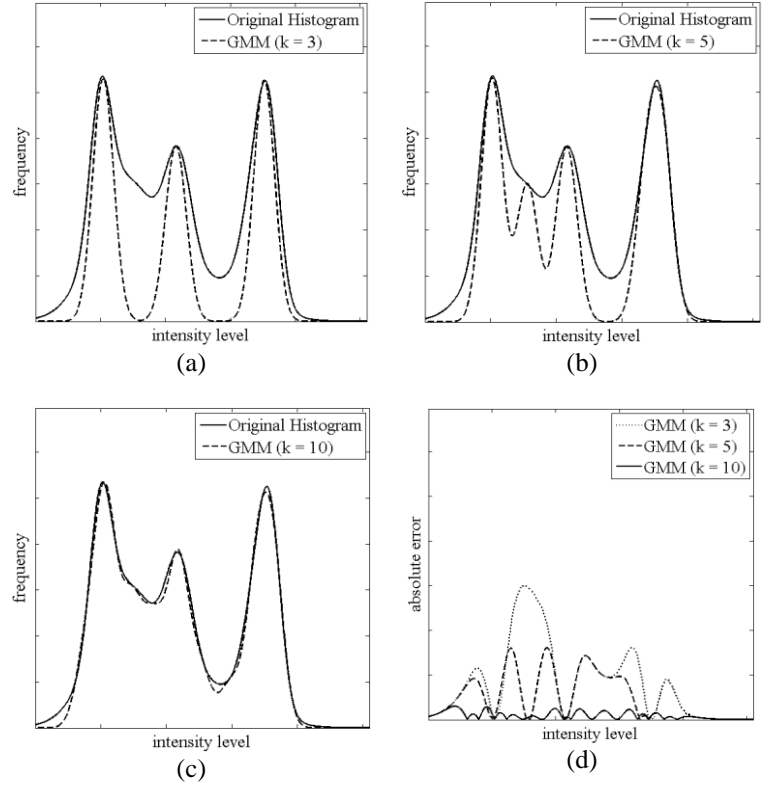

**Fig 2. GMM reconstruction of a histogram by a) 3, b) 5, c) 10 Gaussians and d) their approximation errors**

Therefore, there should be a balance between, using not too few Gaussians to miss the dominant intensity levels and not too many Gaussians to include the non-dominant intensity levels. This is considered in the proposed model by limiting the number of Gaussians to a value where the area under their resultant GMM is no more than *α* percent of the area under the original histogram.

In GMMCE, given a histogram, it searches for an optimal composition of Gaussians to construct the histogram. As an objective function, the optimality of a composition is measured by its similarity to the original histogram. The difference between a GMM and the original histogram is optimized, in a least square sense, as:

$$\arg\min_{\mu,\sigma,\omega} \sum_{I=0}^{L-1} \left( h_{org}(I) - h_{GMM}^k (I \mid \mu, \sigma^2, \omega) \right)^2, \qquad (2)$$

where $h_{org}$ is the original histogram of the image. As discussed earlier, $\mu$, $\sigma$ and $\omega$ are the unknown parameters of the GMM to be estimated.

For optimization one might use the Expectation Maximization (EM) algorithm (like almost all other GMM based optimization problems) which is an iterative approach to estimate the latent variables of a statistical model [20, 22, 26, and 27]. In GMMCE the statistical model is a combination of a specific number of Gaussians and the latent variables include the means and the variances of the Gaussians as well as their scaling factors (i.e. vectors $\mu$, $\sigma$ and $\omega$, respectively).

Although EM has proven to be a very accurate solution for GMM-based problems, it has short falls that make it "not the best approach" for some conditions. The main drawback of EM is its complexity and inability of implementation for real time applications [27]. For example, if the application is auto contrast in digital video recorders [28], it is not feasible to apply EM for every single frame. Moreover, EM-based methods are highly dependent on the starting points. Choosing a bad set of starting points might cause the solution to diverge, which is the case in histogram representation, especially in the presence of abnormal local peaks in the histogram.

*C. Parameter Estimation*

In this section, low complexity greedy approaches are introduced to estimate the abovementioned parameters. In the following sub-sections, three algorithms to be used in GMMCE for parameter estimation of each Gaussian are described. These greedy algorithms are iteratively completed and, at each iteration, the parameters of one Gaussian are estimated and the approximated Gaussian is excluded from the histogram for the estimation of the next Gaussians. These steps are performed until a portion $\alpha$ (e.g. %95) of the underneath area is extracted.

*1) Mean Parameter Estimation*

For mean parameter estimation, given a histogram, the algorithm looks for the intensity level which the sub-histogram around it, has a shape as close to a Gaussian PDF as possible. The algorithm then picks this intensity as the mean of that Gaussian. For this purpose, an objective function is devised that measures this optimality in terms of signal similarity and that finds the best $I \in [0, L-1]$ which satisfies:

$$\hat{\mu} = \arg\max_{I} \sum_{j=0}^{L-1} N(j \mid I, \sigma_0^2) \cdot h_{org}(j) \quad , I = 0, 1, ..., L-1 \qquad (3)$$

where $N(j|I,\sigma_0^2)$ is the value of a normalized Gaussian PDF at $j$, with $I$ and $\sigma_0^2$ as its mean and variance, respectively. Actually, (3) aims at maximizing the correlation (i.e. similarity) between a normalized Gaussian PDF and a region of the original histogram in order to find the best intensity level $I$ where the normalized Gaussian best fits with the original histogram. For any constant value of $\sigma_0^2$ the correlation between a normalized Gaussian and the histogram can locate the position of the largest peak, which is the mean value of that histogram.

One might find mean parameters simply by applying a max operation or using the Dirac delta function instead of $N(j|I,\sigma_0^2)$ in (3). These alternative techniques can be useful only in ideal conditions (i.e. where the shape of the histogram is smooth and local peaks are located exactly on the means of the Gaussians). However, in histograms of natural images it is common to observe distortions in the shape of peak areas, so the output of max operation or Dirac delta function might not accurately locate the corresponding mean parameter. Moreover, since the mean parameter found in this step is the basis for calculations in the following steps, it is essential to find the most accurate value to avoid the error propagation, especially in the variance estimation step.

Implementations show that when the peak value of a Gaussian-shaped sub-histogram is deliberately displaced, the above correlation can still estimate the actual $\hat{\mu}$ based on other frequencies in that neighborhood. This means that compared to Dirac delta function, (3) is more robust to distortions. This robustness is vital for the accuracy of future estimations.

It is also important to note that the selection of the constant value $\sigma_0^2$ is almost independent from the final value of $\hat{\mu}$, since the implementations show that as long as the selected value is not very high, the output will be the same. The reason is that in low contrast images, the histogram is mainly narrow and choosing a small value for variances at this very first step is acceptable.

*2) Gaussian Boundary Analysis*

Before further estimations, it is required to determine the original territory of the ongoing Gaussian-shaped sub-histogram around its estimated mean parameter $\hat{\mu}$. Due to the nature of Gaussian mixture models, each point on the x-axis is actually affected by all Gaussians, proportionally to their tail thickness at that point. In other words, these undesirable effects which exist everywhere on the dynamic range of the histogram are unavoidable when a Gaussian-shaped sub-histogram needs to be merely evaluated. Hence, it is best to find the purest neighborhood around the mean parameter in both directions. Here a neighborhood is considered as pure if the strength of the current Gaussian is dominant compared to both previous and next Gaussians in the histogram.

For this purpose, the proposed algorithm performs a preprocessing filtering to smooth $h_{org}$. First, $h_{org}$ is padded by inserting the adequate number of $h_{org}(0)$ and $h_{org}(L-1)$ in the beginning and the end of histogram, respectively:

$$\begin{cases} h_{org}(-i) = h_{org}(0) \\ h_{org}(255+i) = h_{org}(255) \end{cases}, i = 1, 2, ..., n, \tag{4}$$

Then, a one dimensional averaging filter of size $2 \times n + 1$ is applied on each intensity level to replace each value of the histogram with the average of its neighborhood:

$$h_S(I) = \sum_{j=-n}^{n} h_{org}(j-I) \times \frac{1}{2n+1} \quad , I = 0, 1, ..., 255. \tag{5}$$

After applying the averaging filter, the algorithm sets the lower and upper boundaries (i.e. LB and UB) of the current Gaussian by choosing the nearest local minima of the smoothed histogram (i.e. $h_S$) before and after $\hat{\mu}$, respectively.

*3) Variance Parameter Estimation*

In this section, the estimated mean of a Gaussian-shaped sub-histogram as well as its boundaries (i.e. $\hat{\mu}$, LB and UB) are used as the inputs for the algorithm to estimate the variance. This method brings into play the essential characteristics of a Gaussian PDF and based on the frequency of adjacent intensity levels around the estimated mean, it finds the most probable value for the variance.

Considering a discrete Gaussian PDF with highest similarity to the ongoing Gaussian-shape sub-histogram, one can use the drop ratios of consecutive frequencies in the smoothed histogram to estimate the variance. For instance, equation (6) approximately estimates the drop ratio in the smoothed histogram for the displacement $d$, from peak value at $\hat{\mu}$, toward forward and backward directions ($R_{fw}^d$ and $R_{bw}^d$, respectively):

$$R_{\substack{fw\\bw}}^d = \frac{h_S(\hat{\mu})}{h_S(\hat{\mu} \pm d)} \cong \frac{\omega.N(\hat{\mu}|\hat{\mu},\sigma^2)}{\omega.N(\hat{\mu} \pm d|\hat{\mu},\sigma^2)} = \frac{\omega.\frac{1}{\sqrt{2\pi\sigma^2}}e^{-(\hat{\mu}-\hat{\mu})^2/2\sigma^2}}{\omega.\frac{1}{\sqrt{2\pi\sigma^2}}e^{-(\hat{\mu}\pm d-\hat{\mu})^2/2\sigma^2}} = \frac{\omega.\frac{1}{\sqrt{2\pi\sigma^2}}}{\omega.\frac{1}{\sqrt{2\pi\sigma^2}}.e^{-d^2/2\sigma^2}} = e^{d^2/2\sigma^2}.$$

(6)

Ideally, when the ongoing Gaussian-shaped sub-histogram has the exact same distribution as a discrete Gaussian PDF, only one drop ratio, either in forward or backward direction, is adequate to exactly estimate the variance parameter of every Gaussian-shaped histogram by simply calculating:

$$\hat{\sigma} = \frac{d}{\sqrt{2.\ln(R_{\substack{fw\\bw}}^d)}} \quad , d = 1, 2, ... \tag{7}$$

However in natural images, even after smoothing the original histogram, the histogram is not purely Gaussian so the exact $\hat{\sigma}$ cannot be estimated only with one drop ratio. One solution is to derive it from several ratios around the neighborhood of $\hat{\mu}$ within the lower and upper bounds:

$$\hat{\sigma} = Median(\{\frac{d}{\sqrt{2.\ln(R_{fw}^d)}} \mid d = 1, 2, ..., UB - \hat{\mu}\} \cup \{\frac{d}{\sqrt{2.\ln(R_{bw}^d)}} \mid d = 1, 2, ..., \hat{\mu} - LB\}). \tag{8}$$

From the practical point of view, the wider is the boundary of the Gaussian-shaped sub-histogram, the more probable is the convergence of $\hat{\sigma}$ to its final value. As an example, this method is implemented for the sub-histogram in Fig.1e to estimate the variance parameter for the highest peak of the histogram in the forward direction. Fig.3 indicates that, as the algorithm progresses in the forward direction, the estimated $\hat{\sigma}$ approaches its real value.

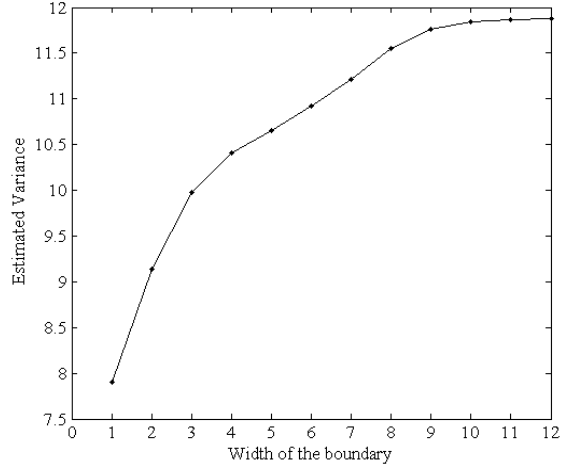

**Fig 3. The convergence of variance estimation algorithm for the sub-histogram in Fig.1e. Seemingly, an acceptable variance would be estimated if more than 7 intensity levels are available within the boundary.**

*4) Scaling Factor Estimation*

After identifying the most dominant sub-histogram by estimating its mean and variance, it should be removed from the global histogram. This can help the next dominant sub-histogram to appear and be identified. For merely considering the ongoing sub-histogram and to avoid affecting other sub-histograms in that neighborhood, the estimated Gaussian PDF can be used. Therefore, the estimated Gaussian PDF is scaled based on the height of the ongoing sub-histogram, and is subtracted from the original histogram:

$$h_S^{t+1}(I) = h_S^t(I) - \hat{\omega}.N(I \mid \hat{\mu}, \hat{\sigma}^2), \qquad I = 0,1,...,L-1 \quad (9)$$

where $h_S^t$ and $h_S^{t+1}$ are the smoothed histograms before and after removal of the current Gaussian-shaped sub-histogram. Now the scaling factor of this sub-histogram $\hat{\omega}$ can be calculated by considering the fact that the subtraction in (9) would modify $h_S^{t+1}(\hat{\mu})$ to zero:

$$h_S^t(\hat{\mu}) - \hat{\omega} N(\hat{\mu} \mid \hat{\mu}, \hat{\sigma}^2) = 0 \Rightarrow \hat{\omega} = \frac{h_S^t(\hat{\mu})}{N(\hat{\mu} \mid \hat{\mu}, \hat{\sigma}^2)} = \frac{h_S^t(\hat{\mu})}{\frac{1}{\sqrt{2\pi\sigma^2}} e^{-\frac{(\hat{\mu}-\hat{\mu})^2}{2\sigma^2}}} = \sqrt{2\pi\sigma^2}.h_S^t(\hat{\mu}).$$

(10)

*D. Intensity Transform Function*

To form the broadened version of the low contrast histogram, the mean parameters are uniformly diffused in the entire range of the histogram (i.e. 0 through *L-1*). First, the dynamic range of the low contrast histogram is set to $[L,U]$ by finding $L$ and $U$, respectively as first and last non-zero significant frequencies (e.g. some sporadic and rare values may be ignored) of the narrow histogram, respectively. Then, the distributed mean values are calculated as:

$$\tilde{\mu}_i = \frac{\hat{\mu}_i - L}{U - L} \times 255 \qquad , i = 1, 2, ..., k. \tag{11}$$

In addition to the mean parameters, the variances of Gaussians are increased by:

$$\tilde{\sigma}_i = \hat{\sigma}_i \times \frac{256}{U - L} \qquad , i = 1, 2, ..., k. \tag{12}$$

With the same scaling factors as those of the narrow histogram (i.e. $\hat{\omega}$), the broadened histogram is formed by a new GMM:

$$H_{GMM}^k (I \mid \tilde{\mu}, \tilde{\sigma}^2, \hat{\omega}) = \sum_{j=1}^{k} \hat{\omega}_j . N(I \mid \tilde{\mu}_j, \tilde{\sigma}_j^2) \qquad , I = 0, 1, ..., L-1 \tag{13}$$

After calculating the destination histogram, a transfer function, that we call *T*, needs to be defined to broaden the narrow histogram. For this purpose, the histogram matching algorithm is implemented. Given two histograms, the source and the destination histograms, histogram matching first calculates their cumulative distribution functions:

$$CDF_{src}(I) = \sum_{j=0}^{I} h_{org}(j) \qquad , I = 0, 1, ..., L-1, \tag{14}$$

$$CDF_{dst}(I) = \sum_{j=0}^{I} H_{GMM}^k (j) \qquad , I = 0, 1, ..., L-1. \tag{15}$$

Then for each intensity level $G_1 \in [0, L-1]$ in the source histogram, it approximates the intensity level $G_2$ in the destination histogram which satisfies $CDF_{src}(G_1) = CDF_{dst}(G_2)$. Finally, the algorithm assigns the approximated intensity level as the output of the histogram matching function *T* for the input $G_1$:

$$T(G_1) = G_2. \tag{16}$$

The enhancement technique of GMMCE is then completed by simply applying the resultant transfer function of *T* to every pixel of the low contrast image.

Considering all steps described in this section, Fig. 4 pictorially demonstrates the whole process of broadening a narrow histogram.

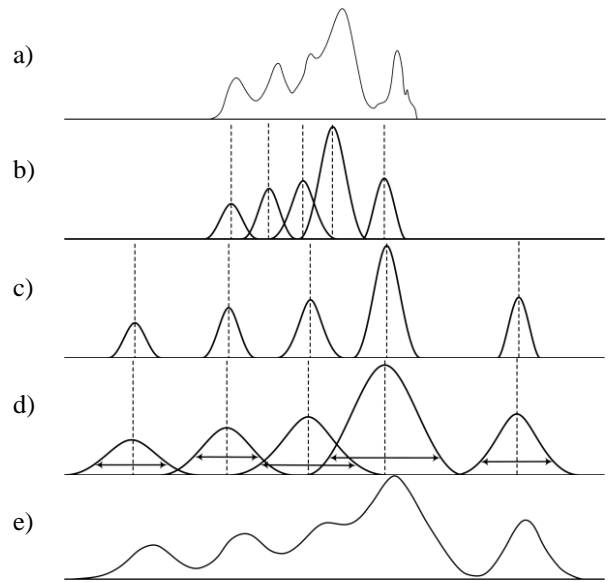

**Fig 4 a) Input narrow histogram of Fig. 1a, b) Estimating 5 Gaussians c) diffusing mean parameters d) stretching Gaussians by increasing variances e) the broadened GMM.**

### III. EXPERIMENTAL RESULTS

To demonstrate the performance of GMMCE, the proposed method was compared with other existing histogram based contrast enhancement methods: Histogram Equalization (HE) [8], Brightness preserving Bi-Histogram Equalization (BBHE) [9], Dualistic Sub-Image Histogram Equalization (DSIHE) [11], Mean-based Recursively Separated and Weighted Histogram Equalization (RSWHE-M with $r=2$ as its best-tuned recursion level) [29], Quadrants Dynamic Histogram Equalization (QDHE) [17]. For thorough comparison, the proposed GMMCE was also compared to a GMM-based method called Expectation Maximization Contrast Enhancement (EMCE) [20]. It can be useful to note that in EMCE all GMM parameters (i.e. means, variances, weights and the number of Gaussians) are estimated by performing the time consuming EM algorithm which makes EMCE a complex contrast enhancement method.

*A. Visual Quality Assessment*

Fig. 5 through Fig. 7 show three low contrast images and the visual quality of their enhanced versions by GMMCE and the other methods. The only parameter used for GMMCE is α = 0.95. Underneath each picture, its histogram is also depicted. For quality comparison, since there is no universally agreed method measuring the quality of contrast enhanced images, we address to their visual quality, as most published works do. Also, note that since histogram based contrast quality algorithms process the histograms of the images, then inspection of their histograms, as shown throughout the pictures, may help in this quality evaluation.

In the *Cars* image (Fig 5) whose original low contrast image has nearly 50 unique intensity levels (i.e. extremely low contrast), it can be seen that the outputs of HE, BBHE and DSIHE is extensively saturated and the output of RSWHE-M is clearly washed-out. However, the results of GMMCE and EMCE can efficiently reveal the details which have been concealed in the original image. Inspections of their histograms also confirm this claim. As seen, the histograms of GMMCE and EMCE are very similar and of the same shape to the original one, reflecting their near identical subjective quality, but are broadened to improve the quality of input picture. For the others, the histogram looks unnatural, especially too much broadened in the vicinity of the main peak. These histograms very well explain the possible artifacts that might appear when shape of an input histogram is not preserved during the enhancements. In this picture, the input narrow histogram has about three dominant peaks and the second one is the most important one. In Fig.5b through Fig5d, the Gaussian-shaped sub-histogram around the most important peak has been extensively modified, so there is almost no mid-gray intensity level remained.

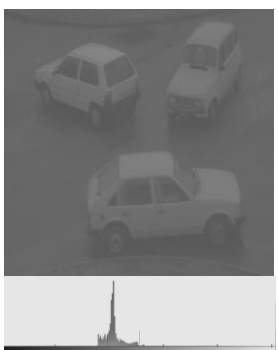 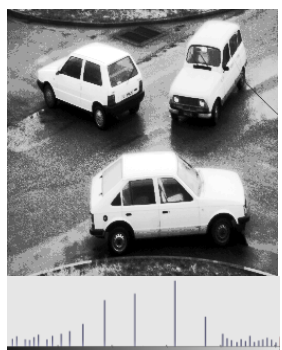 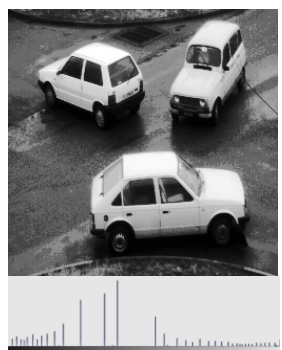 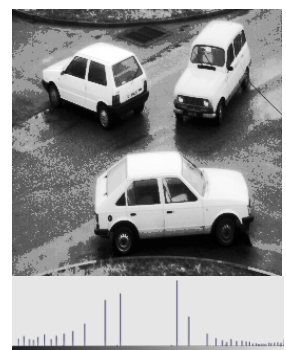

(a) Original    (b) HE    (c) BBHE    (d) DSIHE

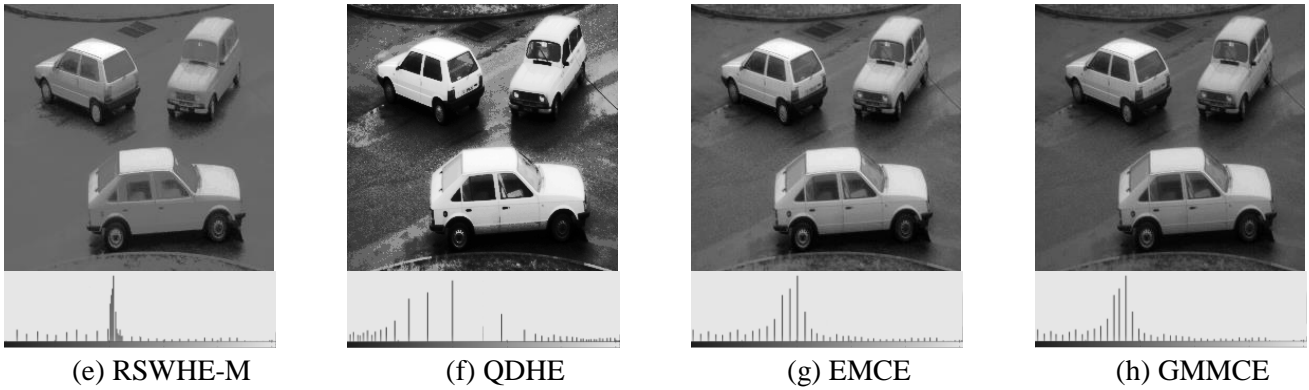

(e) RSWHE-M          (f) QDHE          (g) EMCE          (h) GMMCE

**Fig 5. Contrast enhancement result for image *Cars*.**

In *Cameraman* (Fig.6), the main artifact in the outputs of HE, BBHE, DSIHE, QDHE and RSWHE-M is the non-uniform sky behind the cameraman. This highly bright artifact which is very common in the enhancement of extremely low contrast images with one or more vast uniform regions can be verified in their corresponding histograms. In the original low contrast image, the uniform background related to the second dominant peak (i.e. the wider Gaussian-shaped area in the middle of histogram in Fig.6a which is assumed to be dominant both because of its height and width), is deformed in the enhanced histogram of Fig.6b through Fig.6f. Moreover, in Fig.6e and Fig.6f the enhanced images are also washed-out which are clearly due to the abnormal concentration of intensity levels in the middle of dynamic range of the histograms.

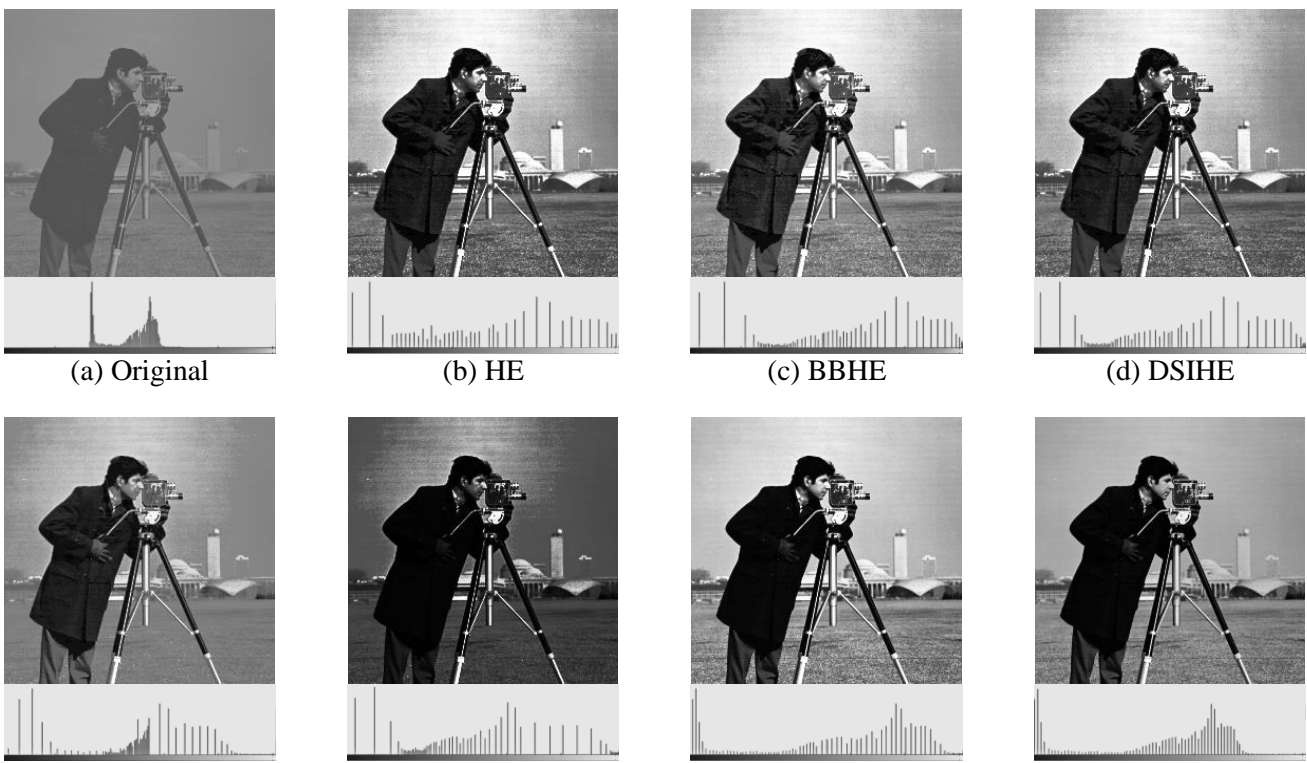

(a) Original          (b) HE          (c) BBHE          (d) DSIHE

| (e) RSWHE-M | (f) QDHE | (g) EMCE | (h) GMMCE |

**Fig 6. Contrast enhancement results for image *Cameraman*.**

Finally, the *Solvay* image (Fig. 7) which is a noisy picture, was enhanced by the abovementioned methods. In this test image, as can be seen from the histogram of Fig. 7a, there are two equally dominant intensity values in the original image which have formed two peaks in the histogram.

Based on the outputs of the methods in Fig. 7, another benefit of histogram shape preservation is demonstrated. For this test image HE, BBHE, DSIHE and QDHE have magnified the natural noise in the original image, by keeping concentrated only the center of the original histogram (noise) and spreading both dominant values over the rest of the dynamic range. Among all of the unsuccessful methods in the previous images, only RSWHE is preserving the shape of the input histogram which has resulted in a satisfying enhanced image, without amplifying the noise. In addition, this image also verifies that EMCE the GMMCE are able to effectively broaden narrow histograms with more than one Gaussian-shaped peak.

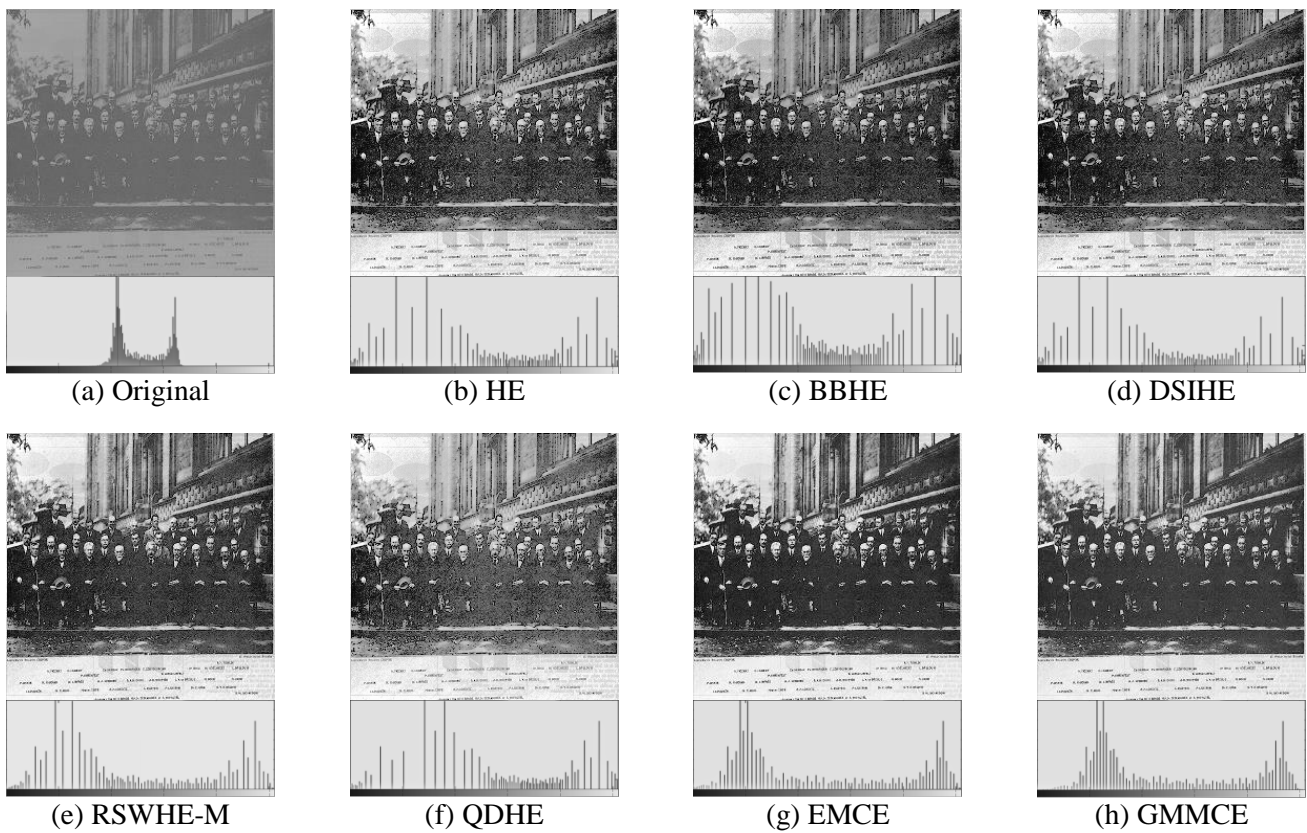

| (a) Original | (b) HE | (c) BBHE | (d) DSIHE |
| (e) RSWHE-M | (f) QDHE | (g) EMCE | (h) GMMCE |

**Fig 7. Contrast enhancement for image *Solvay*.**

*B. Subjective Quality Assessment*

Although the visual quality of GMMCE in preserving the shape of the original image histogram appears to be natural and good, some might argue this is a biased view. People prefer quantitative figures of how good a method can be. This can be achieved by either objective measures or subjective quality scores.

Unfortunately published objective measures are not very reliable. For example, none of the known objective measures consider the saturation effect that might arise from over-enhancing the image. On the other hand the subjective measures should be used with some care, as they normally evaluate the quality of a processed image versus its original quality, which in contrast enhancement becomes meaningless, since the processed image, no matter which method is used, is always better than the original one. In our test, we used an alternative method, where any two enhanced images by two different methods were compared against each other by 19 subjects, In case the assessor could not decisively rate one better than the other, the images shares the same score. Thus when two images are compared, the better one gets a score of 1, the poorer one gets 0 and when they are equally rated, each gets 0.5. In the 7 methods tested here the maximum score one method can get is 6 and the minimum is 0. In addition to the test images in the previous section, three more images shown in Fig. 8 are added. The scores along with the average score of each test image by various methods are tabulated in Table 2. As can be seen the GMMCE along with EMCE, which both have the shape preserving property of the histogram are always rated first or second. On the average EMCE comes slightly above the GMMCE.

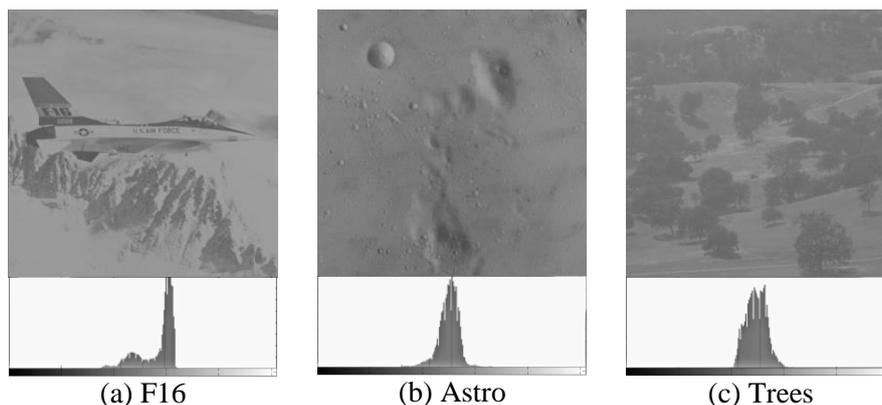

(a) F16        (b) Astro        (c) Trees

**Fig 8. Three additional low contrast images chosen for the subjective quality assessment test**

**Table 2. Comparison of Mean Opinion Score (MOS) of the test images enhanced by the benchmark methods**

| Images | HE | BBHE | DSIHE | RSWHE-M | QDHE | EMCE | GMMCE |
|---|---|---|---|---|---|---|---|
| F16 | 1.07 | 3.21 | 1.85 | 4 | 2.21 | **4.64** | 4.14 |
| Trees | 2.42 | 2.85 | 2.85 | 3.28 | 3.14 | 3 | **3.42** |
| Cameraman | 1.57 | 1.71 | 1.71 | 4.21 | 1.21 | 4.78 | **5.78** |
| Astro | 2.42 | 2.07 | 2.07 | 4.07 | 1.64 | **4.64** | 4.07 |
| Solvay | 1.28 | 2.64 | 1.64 | 4.14 | 0.71 | **5.92** | 4.5 |
| Cars | 0.92 | 3.35 | 0.85 | 3.71 | 1.92 | 4.71 | **5.35** |
| Average | 1.61 | 2.64 | 1.83 | 3.9 | 1.81 | **4.61** | 4.54 |

*C. Computational Complexity Analysis*

In addition to subjective quality assessment, the computational complexities of the abovementioned algorithms are measured. For this purpose, a general-purpose PC with 2.16 GHz Intel® Core 2 Duo processor with 3 MB shared L2 cache and 4 GB of memory is used.

All the seven algorithms were implemented optimally and their average execution times for multiple standard test images were calculated (the used database may be found here: http://decsai.ugr.es/cvg). To avoid any skepticism, all algorithms were run for adequate number of iterations and averages of the execution times are presented in Table 1. For this purpose, each implemented algorithm in MATLAB, was run 60 times for each image. For eliminating the influence of background processing load of the host PC's operating system, executions of the programs were carried out under the same CPU load condition (i.e. monotask). It is also useful to note that, since all methods are histogram-based, their computational loads are mostly related to their processes after calculation of histogram from the RAW files. Thus, results in Table 1 are not grouped by resolution of the test images.

**Table 2. Computational Complexity Measurement**

| Method | Execution time in seconds |
|---|---|
| HE | 0.013 |
| BBHE | 0.166 |
| DSIHE | 0.162 |
| RSWHE-M | 0.128 |
| QDHE | 0.159 |
| EMCE | 0.654 |
| GMMCE | 0.241 |

As discussed earlier, the main reason that EMCE and GMMCE perform well in enhancing very low contrast images, is that they both model histograms by GMMs which is supposed to be a complex process. However, Table 2 shows that replacing EM with greedy algorithms in GMMCE, still places GMMCE among low complexity methods. In contrast, the time consuming EM algorithm in EMCE costs it a major increase of execution time.

## IV. Conclusion

In this paper, a new contrast enhancement method named Gaussian Mixture Model based Contrast Enhancement (GMMCE) has been introduced. First, it is claimed that the shape preservation of narrow histograms during contrast enhancement can avoid unnatural artifacts, such as saturation and wash-out. Based on this claim, the proposed method models the histogram of low contrast image by the combination of a limited number of Gaussians where each Gaussian presents a dominant intensity level of the image. This modeling attempts to reflect the shape of a narrow histogram in the parameters of individual Gaussians, to convey it to a broadened version.

The global contrast enhancement of the image was achieved by the enhancement of sub-histograms separated by the mean value of the Gaussians of the GMM. Experimental results show that the shape preserving method of GMMCE enhances the contrast of the image without generating any undesirable artifact. Moreover, computational complexity analyses show that the greedy parameter estimation algorithms (i.e. instead of time consuming methods like the Expectation Maximization in other GMM based methods) in GMMCE categorize it as a low complexity method and make it suitable for real time applications.

## Acknowledgment

This research was in part supported by a grant from the Institute for Research in Fundamental Sciences (IPM) (No. CS1393-2-04).